\documentstyle[aps,prl,12pt]{revtex}
\input {psfig}

\begin{document}

\title{Strong Pinning and Plastic Deformations of the Vortex Lattice}
\author{A. Sch\"onenberger$^{a}$ \and A. Larkin$^{b,\, c}$ \and E.
  Heeb$^{a}$ \and V. Geshkenbein$^{a,\, b}$ \and G. Blatter$^{a}$}
\address{$^{a\,}$Theoretische Physik, ETH H\"onggerberg, CH-8093
  Z\"urich, Switzerland\\ $^{b\,}$L. D. Landau Institute for
  Theoretical Physics, 117940 Moscow, Russia\\ $^{c\,}$School of
  Physics and Astronomy, University Minnesota, Minneapolis, MN 55455,
  USA}

%\twocolumn[
\date{\today}
\maketitle
%\widetext
%\vspace*{-1.0truecm}
\begin{abstract}
\begin{center}
\parbox{14cm}{
%%%%%%%%%%%%%%%%%%%%%%%%%%%%%%%%%%%%%%%%%%%%%%%%%%%%%%%%%%%%
%                     ABSTRACT
%%%%%%%%%%%%%%%%%%%%%%%%%%%%%%%%%%%%%%%%%%%%%%%%%%%%%%%%%%%%
  We investigate numerically the dynamically generated plastic
  deformations of a 3D vortex lattice (VL) driven through a disorder
  potential with isolated, strong pinning centers (point-like or
  extended along the field direction). We find that the VL exhibits a
  very peculiar dynamical behavior in the plastic flow regime, in
  particular, topological excitations consisting of three or four
  entangled vortices are formed. We determine the critical current density
  $j_c$ and the activation energy for depinning $U_c$ in the presence of a
  finite density of strong pinning centers.

  \medskip\noindent
  PACS numbers: 74.60.-w, 74.60.Ge
}
\end{center}
\end{abstract}
%]
%\narrowtext
\newpage

%\section{Introduction}
%%%%%%%%%%%%%%%%%%%%%%%%%%%%%%%%%%%%%%%%%%%%%%%%%%%%%%%%%%%%%%%%%%%%%%%
Recently, the dynamical behavior of a VL in a disordered type-II
superconductor has attracted much interest
\cite{higgins-bhattacharya-pc-96,koshelev-vinokur-prl-94,%
kwok-fendrich-beek-crabtree-prl-94}.
The intriguing dynamics of the VL originates from the competition
between the vortex -- defect and the vortex -- vortex interaction,
leading to a threshold behavior \cite{campbell-evetts-advph-72}: If
the pinning potential dominates then the vortices are essentially
stationary with a slow residual motion due to tunneling or thermal
activation. In the opposite limit, when the pinning forces are weak
compared with the driving force, the VL is only elastically deformed
and flows coherently.  In the important intermediate regime, when the
pinning and driving force are comparable, the VL deforms plastically
and an incoherent motion results
\cite{shi-berlinsky-prl-91,marley-higgins-bhattacharya-prl-95}.  The
breakdown of the weak collective pinning scenario
\cite{larkin-ovchinnikov-jltp-79} with increasing strength of the
pinning potential has been observed in numerical simulations
\cite{brass-jensen-berlinsky-prb-89}.  Furthermore, it has numerically
been shown \cite{brass-jensen-berlinsky-prb-89,nori-science-96} that
the plastic flow of a 2D VL driven through a random pinning potential
consists of channels of flowing vortices, i.e., {\it rivers}, and
regions of pinned vortex lines, i.e., {\it islands}. This so-called
channel flow behavior has been observed experimentally by means of
Lorentz microscopy
\cite{matsuda-harada-kasai-kamimura-science-96,%
harada-matsuda-bonevich-igarashi-nat-92}.
The various anomalies
\cite{kwok-fendrich-beek-crabtree-prl-94,bhattacharya-higgins-prb-94}
(e.g., thermal instability at large currents, non-monotonicity of the
$I$-$V$ curves, peak effect) occurring in the plastic flow regime
suggest that the nature of the dynamics is fundamentally dissimilar
from the one in the elastic regime. In particular, the dynamically
generated disorder is of general importance as it seems to be a
basic constituent of the dynamics in disordered systems.

In this paper we present for the first time a study on the dynamically
produced defects in a 3D VL. We concentrate on a VL driven through a
material with isolated strong pinning centers. It turns out that such
strong pinning centers can entail the generation of localized
topological excitations involving three or four entangled vortices.
Moreover, the presence of strong pins leads to a preferred orientation
of motion of the VL and affects the critical current density as well as the
barrier for depinning. In the following we first introduce our model
and then consider point-like as well as extended pinning centers.

%
%\section{Model}
%%%%%%%%%%%%%%%%%%%%%%%%%%%%%%%%%%%%%%%%%%%%%%%%%%%%%%%%%%%%%%%%%%%%%
If the density of {\it defects} is low, the VL deformations produced
by individual pinning centers can be studied without accounting for
collective effects. The pinning centers considered here are assumed to
have a rod-like shape with a small lateral (coherence length $\sim
\xi$) but a variable longitudinal (from $\xi$ to a few lattice
constants $a_0$ along the field direction) extension.  This kind of
defects corresponds, for instance, to the discontinuous columns of
damaged material, e.g., YBa$_2$Cu$_3$O$_7$ irradiated with 0.58 GeV Sn
ions \cite{civale-marwick-worthington-kirk-prl-91}, to carbon
nanotubes embedded in Bi$_2$Sr$_2$CaCu$_2$O$_x$
\cite{fossheim-tuset-ebbesen-treacy-pc-95} or to MgO nanorods grown in
BSCCO superconductors \cite{yang-lieber-prpr-96}. The pinning force
with a range of the order of $\xi$ is taken to be infinitely large --
a trapped vortex segment cannot escape from the pinning center -- and,
therefore, only the interaction between the vortices is relevant.
This assumption is physically reasonable since the maximal force the
vortices exert on the pinned vortex segment is much smaller than the
maximal possible force resulting from the depairing current.  The
driving force acting on the vortices is in the plane perpendicular to
the VL and the magnetic field is restricted to $B<0.2 H_{c2}$, i.e.,
the London approximation can be used, and to $\lambda >
a_0=\sqrt{\Phi_0/B}$, where $\lambda$ is the London penetration depth.
These restrictions on the magnetic field are not severe as most of the
experimentally accessible regime is covered.  Within this regime, the
scaling rules \cite{blatter-geshkenbein-larkin-prl-92} are applicable
allowing to generalize the results obtained for isotropic
superconductors considered in the following. For convenience and
technical reasons the moving VL is chosen as the frame of reference
and hence the pinning center appears to move through the VL. In order
to describe the lattice deformation produced by a defect we allow a
finite number of {\it soft} vortices in the neighborhood of the pinned
vortex line to accommodate according to their interaction. All other
vortices are held fixed and consequently denoted as {\it hard}
vortices.  Accordingly, we split the (isotropic) London energy
functional for the soft vortices
\begin{equation}\label{LondonFunctional}
F=\frac{\varepsilon_0}{2}\sum_{i,j}\int d{\bf r}_i \cdot d{\bf r}_j
   \frac{e^{-\sqrt{|{\bf r}_i-{\bf r}_j|}/\lambda}}
        {\sqrt{|{\bf r}_i-{\bf r}_j|}} ,
\end{equation}
into three different parts: The self interaction accounting for the
line energy of the soft vortices, the mutual interaction of the soft
vortices, and the coupling of the soft vortices to the surrounding
hard vortices. The instability to fluctuations of the energy
functional (\ref{LondonFunctional}) is cured by taking a core term
into account and by using a cutoff $\sim \xi$. The algorithm to relax
the vortices is based on a conjugate gradient method; for details on
the numerics, see
Ref.~\cite{schonenberger-geshkenbein-blatter-prl-95}. We do not
investigate the mechanism of vortex trapping and, therefore, the
pinning center considered is initially positioned at a lattice site.
The pinning center fixes the vortex along its length $l_p$.
Subsequently, the pinning center together with the attached vortex
segment is adiabatically dragged through the VL while the deformation
energy of the VL is monitored.  The parameters in this procedure are
the dragging angle $\vartheta$ with respect to a basis lattice vector,
the distance $d$ from the equilibrium lattice site, and the length
$l_p$ of the pinning center. The energy per length of the lattice
deformations is measured in units of $\varepsilon_0$, where
$\varepsilon_0=(\Phi_0/4\pi\lambda)^2$, and the lattice spacing $a_0$
is used as the length unit.

\begin{figure}[htb]
  \centerline{\psfig{figure=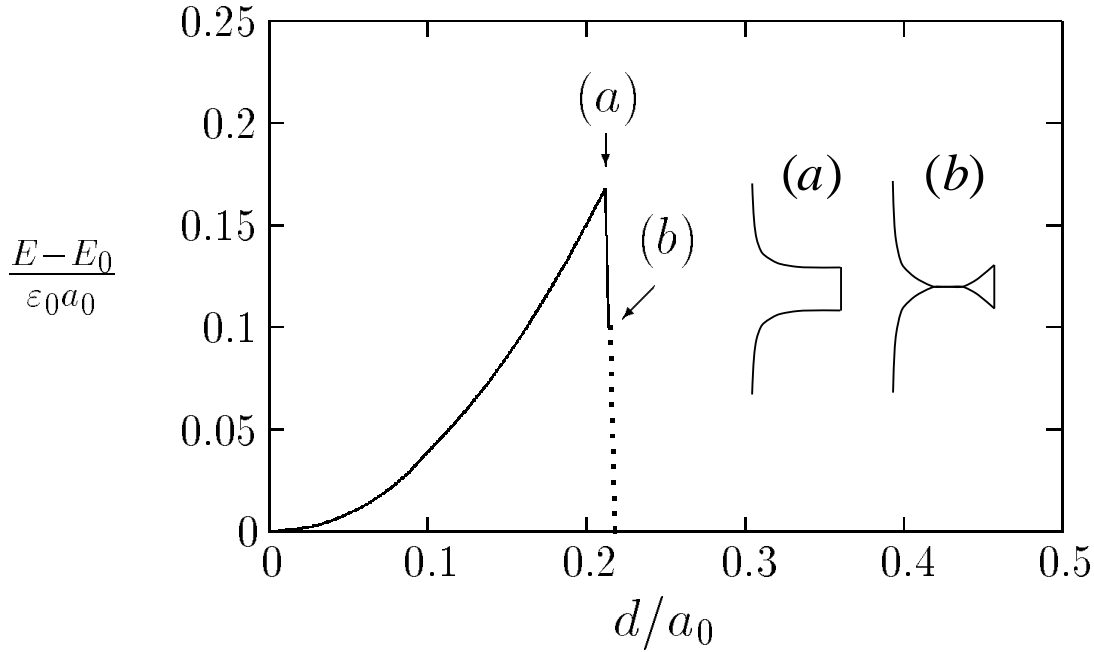,width=0.9\linewidth}}
  \vspace{0.3cm}
  \caption[] 
  {\label{energy_point_pin}The deformation
    energy as a function of the displacement of the pinned vortex
    segment is independent of the angle $\vartheta$ (the curves
    $\vartheta=0^{\circ}$, $15^{\circ}$, and $30^{\circ}$ coincide).
    The magnetic field is fixed at $B\approx 0.0025 H_{c2}$ (i.e.,
    $a_0/\xi=50$) and the pinning length is $3\xi$. The depinning
    occurs at $d_{dp}=0.22 a_0$ and the maximal pinning force
    amounts to $F_{\rm pin}=1.58\varepsilon_0$.}
\end{figure}

%
%\section{Point-like Pinning Centers}
%%%%%%%%%%%%%%%%%%%%%%%%%%%%%%%%%%%%%%%%%%%%%%%%%%%%%%%%%%%%%%%%%%%%%%%
The {\it point-like pinning centers} have an extension of $3\xi$ along
the field direction and the magnetic field $B\approx 0.0025 H_{c2}$ is
fixed by setting $a_0/\xi=50$. Note that the pinning length $l_p=3\xi$
is larger than the cutoff $\xi$ of the energy functional but still
small enough to correspond to point-like pins. Quantitatively similar
results are obtained when $l_p$ is doubled. Estimates on elastic
deformations of the VL suggest that the set of soft vortices should
include the pinned vortex as well as the nearest neighboring vortices.
Small displacements of the pinned vortex segment entail pronounced
deformations only close to the pin where sharps kinks occur.  With
increasing displacement $d$ the parts of the pinned vortex above and
below the pin tend to align anti-parallel and, therefore, attract each
other, see Fig.~\ref{energy_point_pin}. At the critical distance,
$d_{dp}=0.22 a_0$, this attraction dominates and makes the
configuration collapse, i.e., the two anti-parallel vortex segments
annihilate, leaving behind a free unpinned vortex and a vortex loop
attached to the pin. Subsequently, the free vortex line relaxes to its
equilibrium lattice position and the vortex loop shrinks and
disappears. With this {\it one-loop depinning} process the pin is
freed and hence the VL becomes unpinned. Note that the change in
topology is mediated by vortex cutting and recombination.  The
deformation energy versus the displacement is shown in
Fig.~\ref{energy_point_pin}. The sharp edge of the deformation energy
at the recombination of the vortices, see Fig.~\ref{energy_point_pin},
strongly influences the exponent in the scaling law of the depinning
barrier and, in particular, seems to be the signature of depinning
from strong pins.  As the depinning distance $d_{dp}=0.22 a_0$ is
rather small, the pinned vortex line does neither noticeably affect
the other vortices during the dragging process nor experience the
non-circularity of the hexagonal VL potential. Consequently, the whole
dragging process is essentially circular symmetric rendering the
energy of the lattice deformation as well as the depinning mechanism
independent of the angle. As a result, the presence of point-like
pinning centers does not ensue a preferred orientation for a moving VL
and, in addition, no vortex entanglement is introduced.

The critical current density $j_c$ derives from the critical state
and, therefore, is related to the maximal pinning force per defect,
$F_{\rm pin}$, by \cite{ovchinnikov-ivlev-prb-91}
\begin{equation}
j_c=\frac{c\, n\, F_{\rm pin}}{B} .
\end{equation}
Here, $n$ is the volume concentration of effective pinning centers
which depends on the total defect concentration $n_0$ and the
effective trapping area $S$ according to $n=n_0 S/S_0$, where
$S_0=(\sqrt{3}/2)a_0^2$ is the area of the unit cell. On the
basis of the depinning distance $d_{dp}$ an upper limit for $S$ is
obtained by
\begin{equation}
S=\left\{\begin{array}{l}
       \pi\, d_{dp}^2,\quad\;  {\rm if}\;\; \pi\, d_{dp}^2 < S_0, \\
       S_0,\qquad\;\; {\rm otherwise}.
    \end{array}
   \right.
\end{equation}
Inserting the numerical values $F_{\rm pin}=1.58 \varepsilon_0$ and
$d_{dp}=0.22 a_0$ the upper limit for the critical current density
becomes
\begin{equation}
j_c\approx 0.4\,n_0\, a_0^2\,\xi\,j_0 ,
\end{equation}
where $j_0=(4/3\sqrt{3})(c\varepsilon_0)/(\xi\Phi_0)$ is
the depairing current density. Note that $j_c$ is inversely proportional to
$B$. For the creep activation energy, $U(j)$, we obtain
\begin{equation}\label{U_jc}
 U(j)=U_c \left(\frac{j_c-j}{j_c}\right)^{\alpha} ,
\end{equation}
with $U_c=0.18 \pm 0.02 \varepsilon_0 a_0$ and $\alpha\approx 1.97$.
The exponent $\alpha\approx 2$, resulting from the sharp drop in
energy at depinning, differs from the value $5/4$ obtained
for a string in a (smooth) washboard pinning potential
\cite{blatter-feigelman-geshkenbein-larkin-rmp-94}.

%
%\section{Extended Pinning Centers}
%%%%%%%%%%%%%%%%%%%%%%%%%%%%%%%%%%%%%%%%%%%%%%%%%%%%%%%%%%%%%%%%%%
Elastic considerations suggest that {\it extended pinning centers}
deform not only the nearest neighboring vortices but also some of the
next nearest neighbors and, therefore, all these vortices must be
soft. In order to study the effect of extended pins, the magnetic
field is fixed at $B\approx 0.06H_{c2}$ corresponding to $a_0/\xi=10$
and the pinning centers are restricted to lengths $l_p \le 7 a_0$.
Firstly, we analyze the topological aspects of the lattice
deformations and, secondly, we determine the critical current density
and the activation energy for depinning.

\begin{figure}[h]
\noindent
 \hspace{0.1\linewidth}
 \begin{minipage}[b]{.22\linewidth}
  \centerline{\psfig{figure=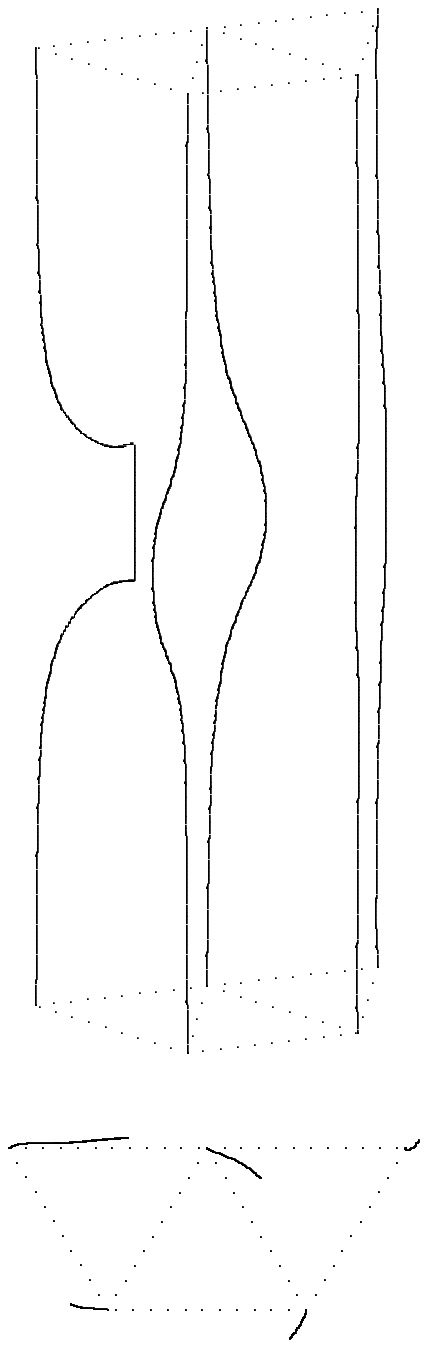,width=\linewidth}}%
\hspace{0.5cm}$(a)$
 \end{minipage}
 \begin{minipage}[b]{.22\linewidth}
  \centerline{\psfig{figure=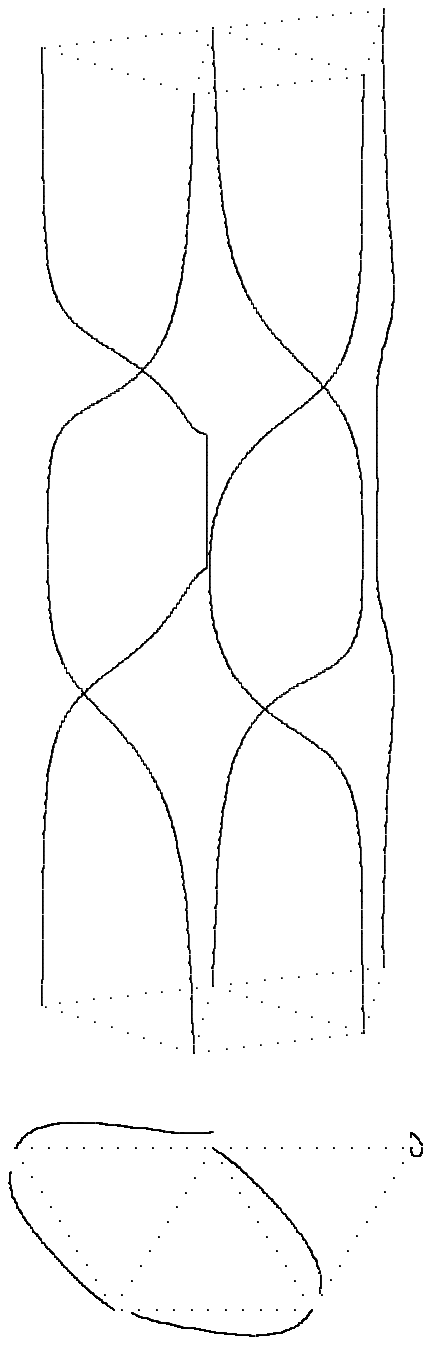,width=\linewidth}}%
\hspace{0.5cm}$(b)$
 \end{minipage}
 \begin{minipage}[b]{.22\linewidth}
  \centerline{\psfig{figure=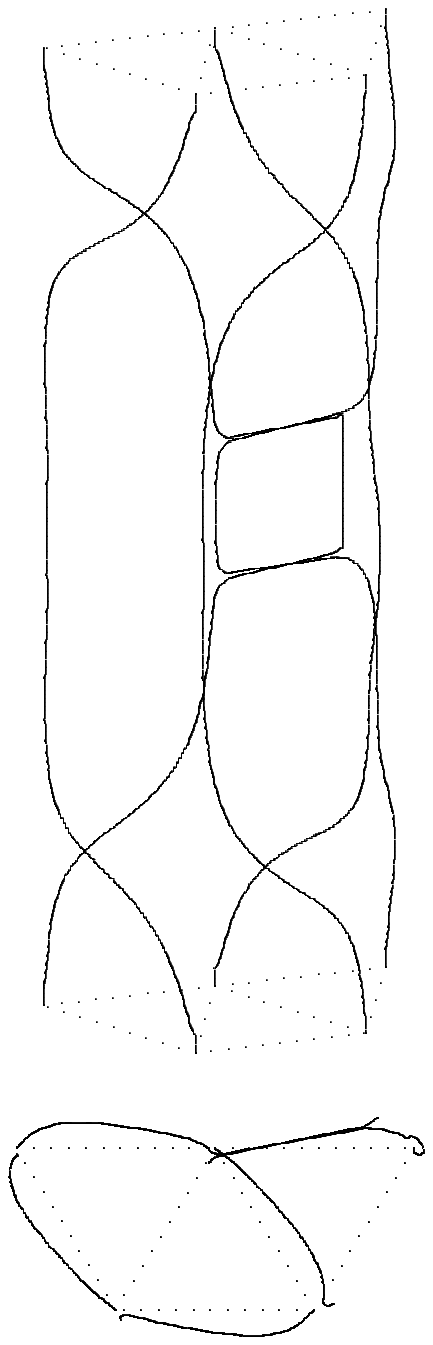,width=\linewidth}}%
\hspace{0.5cm}$(c)$
 \end{minipage}
 \hspace{0.1\linewidth}
 \vspace{0.3cm}
 \caption[]
 {\label{TQ_deform}An extended pinning center with $l_p=3 a_0$
   together with the pinned vortex segment is dragged through the VL
   at an angle $\vartheta=5^{\circ}$. Two TQ's appear at $=0.98 a_0$
   and disappear again after the vortex has depinned via a pair
   annihilation state.}
\end{figure}

\begin{figure}[h]
  \centerline{\psfig{figure=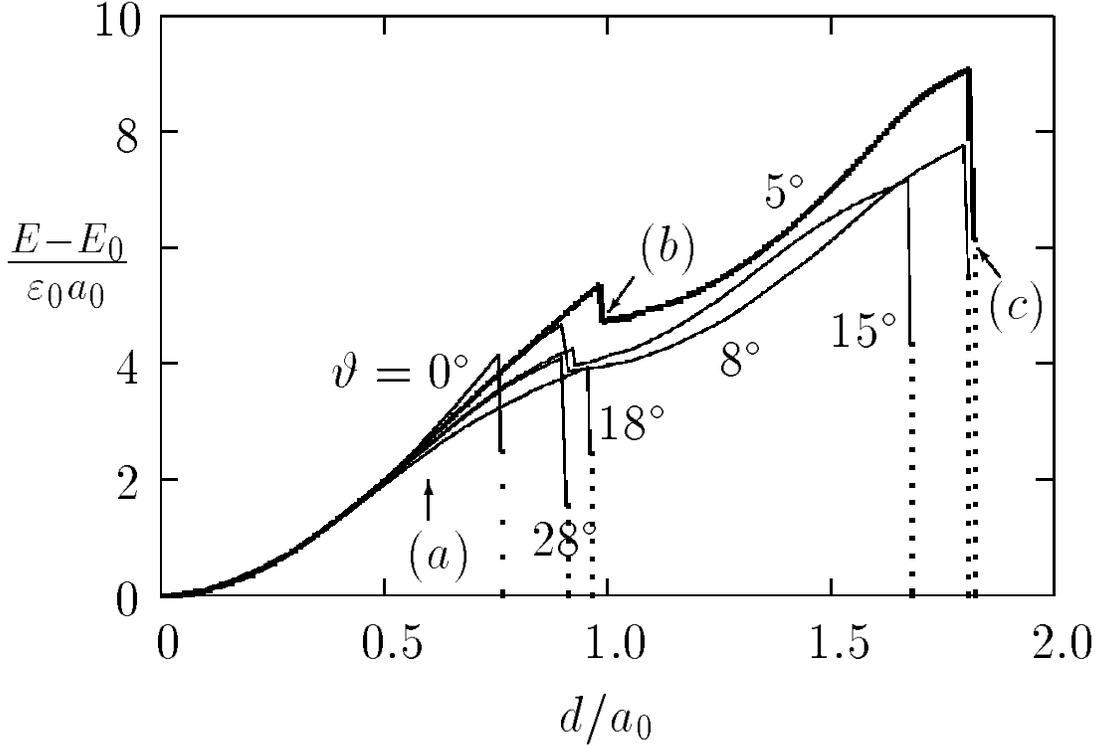,width=0.9\linewidth}}
  \vspace{0.3cm}
  \caption[]
  {\label{Energy_TQ}The deformation energy exhibits a palpable drop
    when the topology changes, i.e., for a displacement $d\sim a_0$.
    For $\vartheta=0^{\circ}$ a one-loop depinning occurs whereas for
    small angles, e.g., $\vartheta=5^{\circ}$, TQs are generated.
    Intermediate angles, e.g., $\vartheta=8^{\circ}$ and
    $\vartheta=15^{\circ}$ lead to the creation of TTs.  For larger
    angles $\vartheta=18^{\circ}$ and $\vartheta=28^{\circ}$ a pair
    annihilation state is formed. Note that the largest slope of the
    energy curve, obtained for $\vartheta=0^{\circ}$, determines the
    maximal pinning force and hence the critical current density as
    well as the orientation of motion.  The configurations $(a)$,
    $(b)$, and $(c)$ are displayed in Fig.~\ref{TQ_deform}.}
\end{figure}

For extended pinning centers the situation is more complicated than
for point pins and the lattice deformations produced depend crucially
on the angle $\vartheta$, the pinning length $\l_p$, and the distance
$d$. In order to illustrate the effects of extended pins we consider a
pin with a length $l_p=3 a_0$ dragged along an angle
$\vartheta=5^{\circ}$.  At a distance $d=0.98 a_0$ configurations of
four twisted vortices, so-called {\it twisted quadruplets} (TQ), are
formed above and below the pinning center, i.e., a loop and anti-loop
excitation of four vortices, see Fig.~\ref{TQ_deform}.
\noindent
At the transition from the disentangled to the entangled state the
deformation energy drops considerably $\sim (0.5 - 1) \varepsilon_0
a_0$, see Fig.~\ref{Energy_TQ}, and becomes slightly larger than the
energy of two TQs because the pin is not at a lattice site.  The high
energy of these loop excitations, $E_{TQ}-E_0=2.41 \varepsilon_0 a_0$
($E_0$ is the ground state energy of the regular VL), is due to the
hexagonal VL which confines the TQ to a length $l_{TQ}=2.94 a_0$. The
TQs above and below the pinning center have opposite orientation and,
therefore, attract each other.  The interaction energy of the TQs can
be estimated by considering two oppositely oriented vortex rings in
parallel planes separated by a distance $L$.  Choosing vortex rings on
top of each other with the same radius $R$, we obtain
\begin{equation}
E_L=\pm 2\pi^2\frac{R^4}{L^3}\varepsilon_0
     +O\left(\frac{R^6}{L^5}\varepsilon_0\right) .
\end{equation}
Using the numerically obtained values $L\approx 6.4 a_0$ and $R\approx
a_0$ we find
\begin{equation}
E_L(TQ)\approx -2\pi^2\varepsilon_0 a_0^4/L^3
       \approx -0.08\varepsilon_0 a_0 .
\end{equation}
When the pinned vortex segment is at its closest distance from its
nearest neighboring vortex, the configuration is similar to the
initial state since the TQs ``carry away'' the displacement of one
lattice constant.  Therefore, it looks as if the nearest neighbor
rather than the original vortex were pinned. 
If the pinned vortex segment is dragged beyond the nearest neighboring
vortex towards the next nearest neighbor the distance between the TQs
is increased. Despite the larger separation of the TQs, there is not
enough space for the creation of a second pair of TQs (this behavior
has been obtained for $l_p\le 7 a_0$). Instead, the pinned vortex and
the next nearest neighboring vortex form a configuration of two
twisted vortices. Since configurations of two twisted vortices are
unstable in the VL \cite{schonenberger-geshkenbein-blatter-prl-95} the
two vortices collapse and a {\it pair annihilation state} is formed,
leading to an exchange of the pinned vortex segment, see
Fig.~\ref{TQ_deform}. After depinning the two oppositely oriented TQs
move together and annihilate, leaving behind a ``healed'' regular VL.
The depinning terminates the generation of vortex entanglement
associated with the formerly pinned vortex. However, configurations of
entangled vortices can be stabilized by other pins, in particular, weak
point disorder.

\begin{figure}[htb]
 \centerline{\psfig{figure=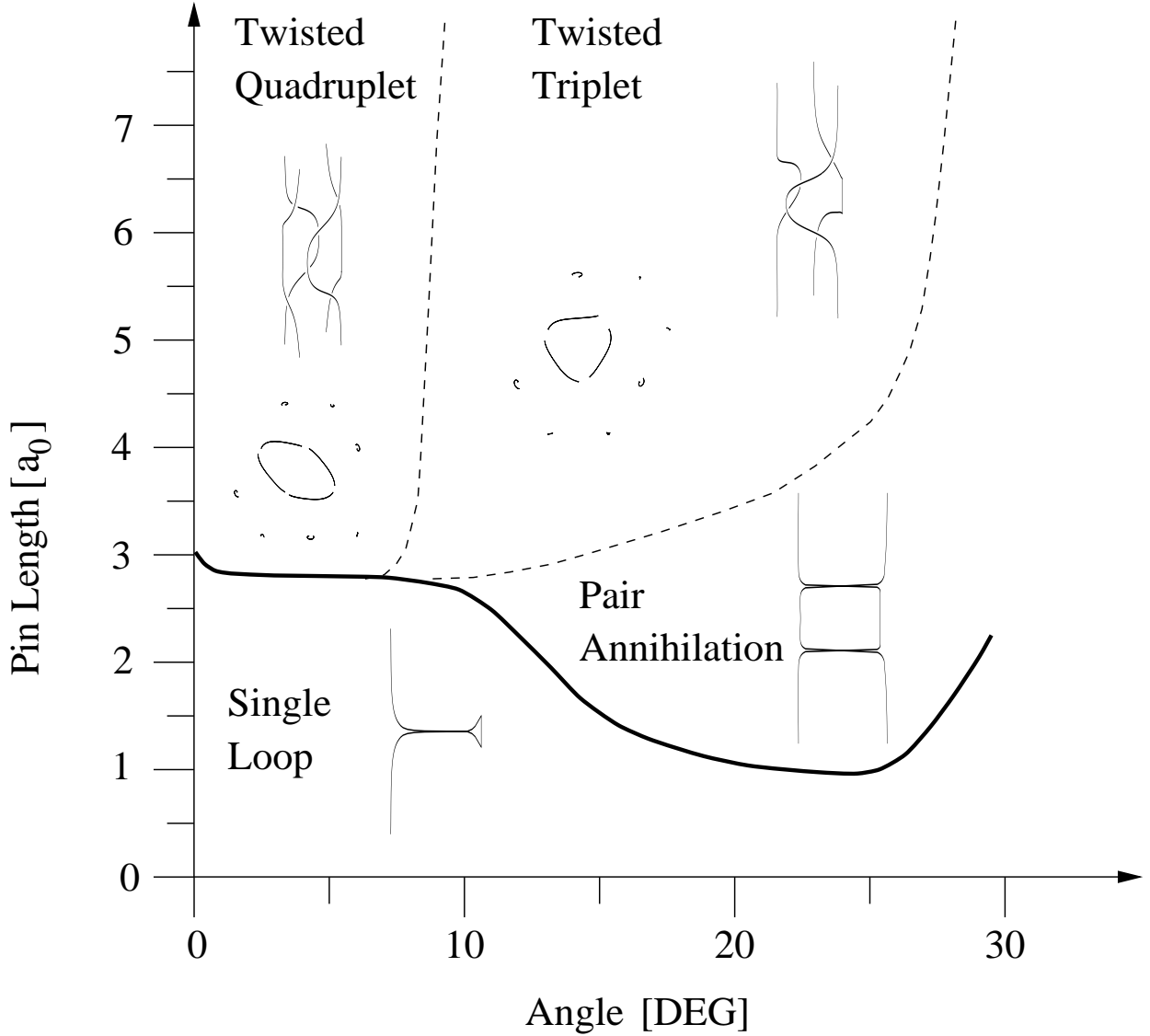,width=\linewidth}}
 \vspace{0.3cm}
 \caption[]
 {\label{deform_diagram}Extended pinning centers can trigger the
   formation of configurations with three or four twisted vortices.
   Below the solid line depinning is mediated by a vortex loop
   creation at the pin, whereas above this line depinning always
   involves pair annihilation either with the nearest or the next
   nearest neighboring vortex.}
\end{figure}

The formation of two TQs is a generic case for extended pinning
centers. With increasing angle, the force the pinned vortex
exerts onto its closest neighboring vortex is reduced and, therefore,
only configurations of three twisted vortices, so-called {\it twisted
  triplets} (TT), are generated. The TTs behave similarly to the TQs
and, in particular, disappear in the same way. For angles close to
$\pi/6$ a pair annihilation state is formed (note that the twisted
pair is unstable in the VL) which in turn leads to an exchange of the
pinned vortex segment with the nearest neighboring vortex.  The
lattice deformations as a function of the angle $\vartheta$ and the
pinning length $l_p$ are summarized in the ``phase diagram''
in Fig.~\ref{deform_diagram}.
The strong dependence of the deformation energy on the angle
$\vartheta$, see Fig.~\ref{Energy_TQ}, makes the VL move along a
lattice vector. This behavior is a consequence of the maximal pinning
force which is obtained for $\vartheta=0^{\circ}$, see
Fig.~\ref{Energy_TQ}.  Interestingly, weak pinning seems to introduce
the same orientation of motion as has been demonstrated by experiments
on Al films \cite{fiory-prl-71} as well as by theoretical studies
\cite{schmid-hauger-jltp-73,giamarchi-doussal-prl-96}. The critical
current density for extended pins derives along the same lines as for
point pins, but with $S=S_0$, as the change in topology occurs at
$d\sim a_0$.  Accordingly, the critical current density is given by
\begin{equation}
 j_c=\frac{3\sqrt{3}}{4}\frac{F_{\rm pin}}{\varepsilon_0}
     n_0\,a_0^2\,\xi\,j_0 .
\end{equation}
$F_{\rm pin}$ depends on the field and, therefore, the critical
current density is not inversely proportional to $B$ (as is the case
for point pins) but is given by
\begin{equation}
 j_c\approx 6.1 \left(\frac{B_{lp}}{B}\right)^{\gamma}n_0\,\xi\,l_p^2\,j_0 ,
\end{equation}
with $\gamma=0.7\pm 0.1$, where $B=\Phi_0/a_0^2$ is normalized with
respect to $B_{lp}=\Phi_0/l_p^2$. The activation energy to change the
topology, however, exhibits the same scaling behavior as for
point-pins
\begin{equation}
  U(j)=U_c(l_p)\left(\frac{j_c-j}{j_c}\right)^{\alpha} ,
\end{equation}
with $\alpha\approx 2$, $U_c(l_p)=\beta \varepsilon_0 a_0
(l_p/a_0)^{\gamma}$, $\beta=1.63\pm 0.05$ and $\gamma=0.78\pm 0.05$. The
universality of the exponent $\alpha$ originates from the pronounced
edge of the deformation energy when the topology changes. 

In conclusion, we have studied the plastic deformations of the vortex
lines due to strong pinning. We have identified the relevant depinning
processes involving either loop creation or pair-annihilation,
depending on the pin size. We have found that large pins do generate
entangled configurations (TTs and TQs) which will be stabilized by a
finite density of pins. Finally, we have determined the critical
current density $j_c$ as well as the activation barrier $U(j)$ for
creep, the latter showing a universal exponent $\alpha\approx 2$
originating from the change in topology at depinning.

We gratefully acknowledge the financial support from the Swiss
National Science Foundation.

%\bibliographystyle{aip}
%\bibliography{journals,ReferencesPaper}

\end{document}